\documentclass[iop]{emulateapj} 

\usepackage{apjfonts,natbib,pstricks,pst-grad,amsmath,booktabs,color,epsfig}

\shortauthors{Castro \& Slane}
\slugcomment{Accepted for publication in ApJ, 2010 May 11}

\begin{document} 

\title{{\it FERMI} LAT OBSERVATIONS OF SUPERNOVA REMNANTS INTERACTING WITH MOLECULAR CLOUDS}
\author{Daniel Castro\altaffilmark{1,2}, and Patrick Slane\altaffilmark{1}}

\altaffiltext{1}{Harvard-Smithsonian Center for Astrophysics, 60 Garden Street, Cambridge, MA 02138, USA}
\altaffiltext{2}{Departamento de F\'isica, Universidad Sim\'{o}n Bol{\'\i}var, Valle de Sartenejas, Apdo. 89000, Caracas 1080A, Venezuela}

\begin{abstract}
We report the detection of $\gamma$-ray emission coincident with four supernova remnants (SNRs) using data from the Large Area Telescope on board the {\it Fermi Gamma-ray Space Telescope}. G349.7+0.2, CTB 37A, 3C 391, and G8.7--0.1 are SNRs known to be interacting with molecular clouds, as evidenced by observations of hydroxyl (OH) maser emission at 1720 MHz in their directions. SNR shocks are expected to be sites of cosmic ray acceleration, and clouds of dense material can provide effective targets for production of $\gamma$-rays from $\pi^0$-decay. The observations reveal unresolved sources in the direction of G349.7+0.2, CTB 37A, and 3C 391, and a possibly extended source coincident with G8.7--0.1, all with significance levels greater than 10$\sigma$. 

\end{abstract}


\section{INTRODUCTION}

The case for cosmic ray acceleration at supernova remnant (SNR) shocks has found increasing support in observational evidence, both from the thermal and non-thermal components of the emission from these objects. The production of relativistic electrons has been clearly established from the non-thermal X-ray emission detected from several shell-type SNRs, e.g., SN 1006 \citep{Koyama1995,Reynolds1998}, RX J1713.7-3946 \citep{Koyama1997,Slane1999}, and Vela Jr. \citep{Aschenbach1998,Slane2001}. Additionally, the dynamical properties of some SNRs suggest that a significant fraction of the explosion energy is going into the acceleration of ions to relativistic speeds, like in the case of 1E 0102.2-7219 \citep{Hughes2000}. X-ray measurements of the relative positions of the forward and reverse shocks, and
contact discontinuity in Tycho's SNR also indicate modification of the remnant dynamics by the acceleration of cosmic-ray ions \citep{Warren2005}. $\gamma$-ray emission in the TeV range from some SNRs, observed with ground-based telescopes, seems to confirm that particles are indeed being accelerated to energies approaching the {\it knee} of the cosmic ray spectrum \citep{Enomoto2002,Aharonian2007b,Katagiri2005,Aharonian2007a}. However, there is growing debate as to whether the origin of this high energy emission is hadronic or leptonic in nature, as reviewed in \citet{Reynolds2008}.
	
Pion production, followed by $\pi^0\!\! \rightarrow\!\! \gamma \gamma$ decay, is a natural result of the interaction of relativistic ions, like those generated through diffusive shock acceleration at SNR shocks, with dense material \citep{Drury1994, Baring1999}. Hence, SNRs known to be interacting with molecular clouds present particularly interesting targets for the detection of $\gamma$-ray emission of hadronic origin, as suggested by \citet{Frail1996} and \citet{Claussen1997}. The detection of hydroxyl (OH) masers at 1720 MHz is a very useful tool for finding this type of interactions, as they indicate the presence of shocked clouds of dense molecular material \citep{Wardle2002,Hewitt2009b}. 

Observations in the MeV to GeV range are especially important in order to identify signatures of $\pi^0$-decay, because the photon spectrum from this process has a distribution that peaks approximately at 70 MeV \citep{Reynolds2008}. The Energetic Gamma-Ray Experiment Telescope (EGRET), on board the Compton Gamma-Ray Observatory, found a host of discrete $\gamma$-ray sources in the GeV band possibly associated with Galactic SNRs \citep{Torres2003}. However, due to poor spatial resolution and other instrumental constraints, it has proven impossible to determine unambiguously the origins of these EGRET sources. The Large Area Telescope (LAT), onboard the {\it Fermi Gamma-ray Telescope} has opened a new window for studying SNRs in the GeV energy range, due to its improved sensitivity and resolution. For example, the detection of an extended LAT source coincident with SNR W51C (G49.2-0.7), was reported by \citep{Abdo2009a}, and their study suggests that $\pi^0$-decay emission is the dominant component of the $\gamma$-ray flux observed in its direction. The observation of OH maser (1720 MHz) emission in the direction of this SNR, reported by \citet{Green1997}, suggests that it is interacting with a cloud of dense material.

SNRs G349.7+0.2, CTB 37A, 3C 391 and G8.7--0.1, are known to be interacting with molecular clouds, as evidenced by OH masers observed in their directions \citep{Frail1996,Hewitt2009a}. In this paper we report the detection of four LAT sources coincident with these SNRs, and on the results of our detailed study of $\gamma$-ray emission in their regions. After a description of how the LAT data are analyzed in this work in Section 2, we present the results of the observations and put them in context for each individual object in Section 3. A discussion of the results is given in Section 4.

\section{OBSERVATIONS AND DATA TREATMENT}

The LAT $\gamma$-ray detector is the main instrument on board the {\it Fermi Gamma-ray Space Telescope}, which was launched on 2008 June 11. The LAT is a pair-conversion, imaging, large field-of-view (FoV) $\gamma$-ray telescope, that covers a wide energy range, from 20 MeV and up to 300 GeV. High precision tracking of electrons and positrons resulting from pair conversion allow the LAT to reconstruct the direction of $\gamma$-rays incident on the detector, and measurements of the energy of the subsequent electromagnetic showers are performed in its calorimeter. The FoV is $\sim$ 2.3 sr and the point-spread function (PSF) varies largely with photon energy, and improves at higher energies. The converter-tracker device has 16 planes of high-{\emph Z} material, which are divided in two regions. The {\it front} section contains the first 12 layers which are thinner and optimize the PSF, and the {\it back} region has 4 thicker planes, which maximize the effective area at the expense of a lower angular resolution relative to the {\it front} section. Further details about the instrument and data calibration are described in \citet{Atwood2009}.

The LAT, while operating on "sky survey" mode, observes the entire sky every two orbits (approximately 3 hr). In this work, 15 months of LAT data (from 2008 August until 2009 November) are analyzed. Only events belonging to the {\it diffuse} class, which reduces the residual background rate as explained in detail in \citet{Atwood2009}, have been selected for this study. The updated instrument response functions (IRFs) used are called "Pass6 version 3", which were developed using in-flight data and consider pile-up and coincidence effects not included in the pre-launch analysis \citep{Rando2009}. The systematic uncertainties of the effective area, for the IRF used, are energy dependent: 10\% at 100 MeV, decreasing to 5\% at 560 MeV, and increasing to 20\% at 10 GeV \citep{Porter2009}. Additionally, only events coming from zenith angles smaller than 105$^{\circ}$ are selected in order to reduce the contribution from terrestrial albedo $\gamma$-rays \citep{AbdoPRD}.

The  $\gamma$-ray data in the direction of G349.7+0.2, CTB 37A, 3C 391 and G8.7--0.1 are analyzed using the Fermi Science Tools\footnote{The Science Tools package v9r15p2, and related documentation, is distributed by the Fermi Science Support Center at http://fermi.gsfc.nasa.gov/ssc}. In order to study spatial distribution, position and spectral characteristics, the maximum likelihood fitting technique is employed using {\it gtlike}. The likelihood statistical approach is used on LAT data because of the low detection rates and the extent of the PSF, and it estimates the best model parameters by maximizing the joint probability for the data given the emission model \citep{Mattox1996}. The emission models used in {\it gtlike} place individual sources, with particular spectral properties, at fixed positions, and include a Galactic diffuse component resulting from the interactions of cosmic rays  with interstellar material and photons, and an isotropic one that accounts for the extragalactic diffuse and residual backgrounds. In this work, the mapcube file gll\_iem\_v02.fit is used to describe the $\gamma$-ray emission from the Milky Way, and the isotropic component is modeled using the isotropic\_iem\_v02.txt table. 

\subsection{Spatial Analysis}

Spatial analysis is performed using $\gamma$-ray data in the energy range 2-200 GeV, and converted in the {\it front} section, in order to improve the angular resolution. The 68\% containment radius angle for normal incidence {\it front}-selected photons in this energy band is $\leq 0^\circ.3$. To analyze the neighborhoods of the SNRs studied, count maps of the $10^\circ \times 10^\circ$ region centered on the region of interest (ROI) are constructed, and smoothed by a Gaussian kernel to match the PSF of the events selected.
 
In order to determine the detection significance and the precise position of the sources associated with the objects of interest, Galactic and isotropic backgrounds are modeled and test statistic maps are constructed using {\it gttsmap}.  The test statistic is the logarithmic ratio of the likelihood of a point source being at a given position in a grid, to the likelihood of the model without the additional source, $2\text{log}(L_{\text{ps}}/L_{\text{null}})$. The resolution of these maps, set by the size of the grid used, is $0^\circ.05$. 

\subsection{Spectroscopy}

The study of the spectral energy distribution (SED) characteristics of each of the sources associated with G349.7+0.2, CTB 37A, 3C 391, and G8.7--0.1, is performed through likelihood analysis with {\it gtlike}. Events converted in both {\it front} and {\it back} sections, and in the energy range 0.2-51.2 GeV, are included in the study. The lower energy bound is selected to avoid source confusion, to avoid the rapidly changing effective area of the instrument at low energies, and because of the large uncertainty below 0.2 GeV related to the Galactic diffuse model used. 

The spectral analysis is performed using {\it gtlike} to model the flux at each energy bin and estimate, through the maximum likelihood technique, the best-fit parameters. Background sources from the 3 month {\it Fermi} LAT Bright Gamma-ray Source List \citep{Abdo2009c}, and detected pulsars from the 11 month {\it Fermi} LAT Pulsar Catalogue \citep{Abdo2009b}, within a $10^\circ$ radius of the sources of interest are modeled. Sources detected in the test statistic map of each field with significance greater than $5\sigma$ are also included in the background model. Additionally, the analysis was repeated using all background sources in the recently released one-year {\it Fermi} LAT First Source Catalog (1FGL) in each region\footnote{ The data for the 1451 sources in the {\it Fermi} LAT First Source Catalog is made available by the Fermi Science Support Center at http://fermi.gsfc.nasa.gov/ssc/data/access/lat/1yr\_catalog/}, with no significant effect on the results. The normalization of the Galactic diffuse component is left free at each energy step. As an addition to the statistical uncertainties associated with the likelihood approach, and the systematic errors related to the IRF as mentioned above, the uncertainty of the underlying Galactic diffuse level is considered. The treatment of this error in the evaluation of the systematic uncertainties is performed by artificially changing the normalization of the Galactic background by $\pm3\%$, similarly to the analysis in \citet{Abdo2009a}.

\section{RESULTS AND ANALYSIS}

\subsection{G349.7+0.2}

\begin{center}
\begin{figure*}
\begin{center}
\includegraphics[width=0.45\textwidth]{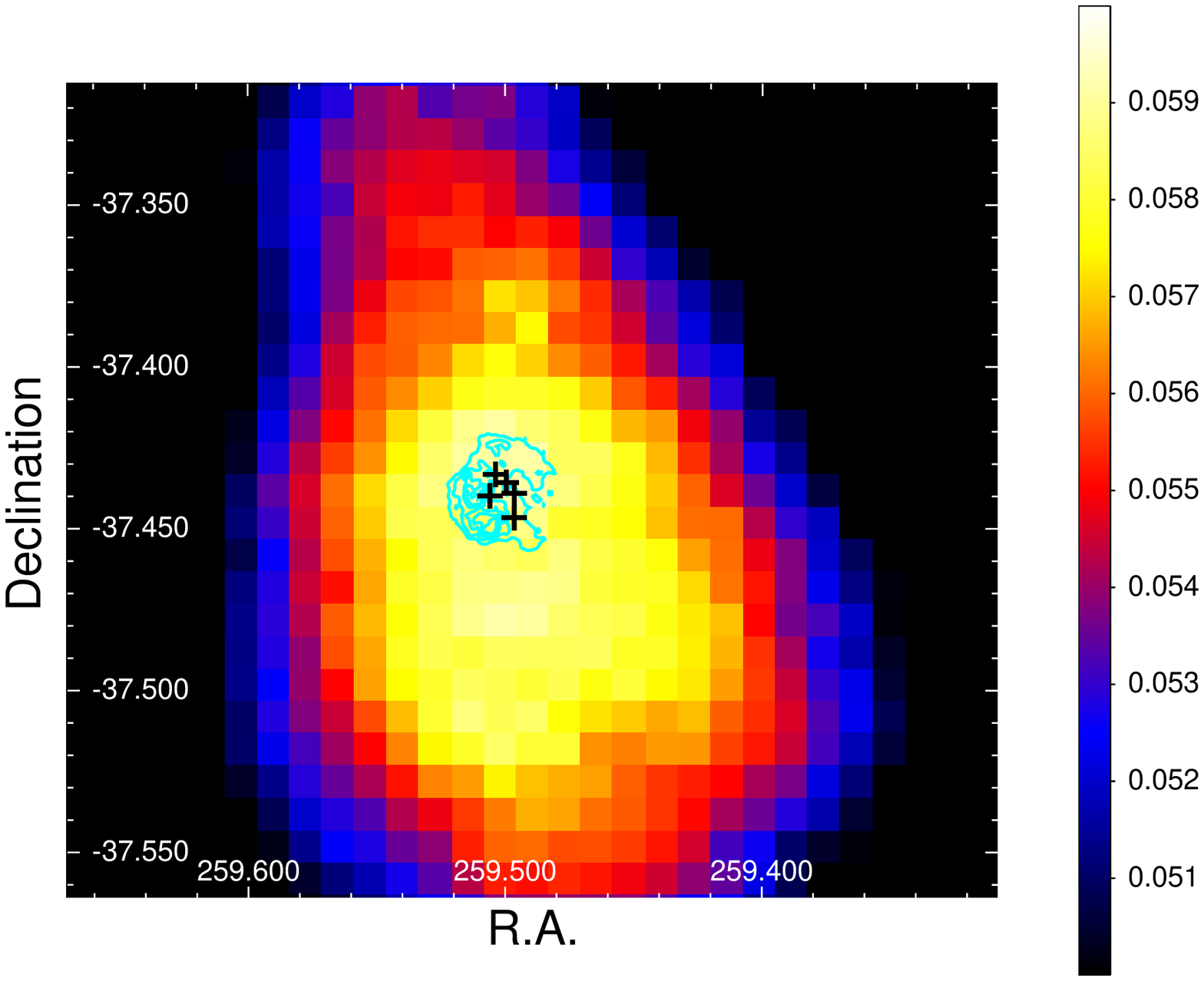}  
\includegraphics[width=0.45\textwidth]{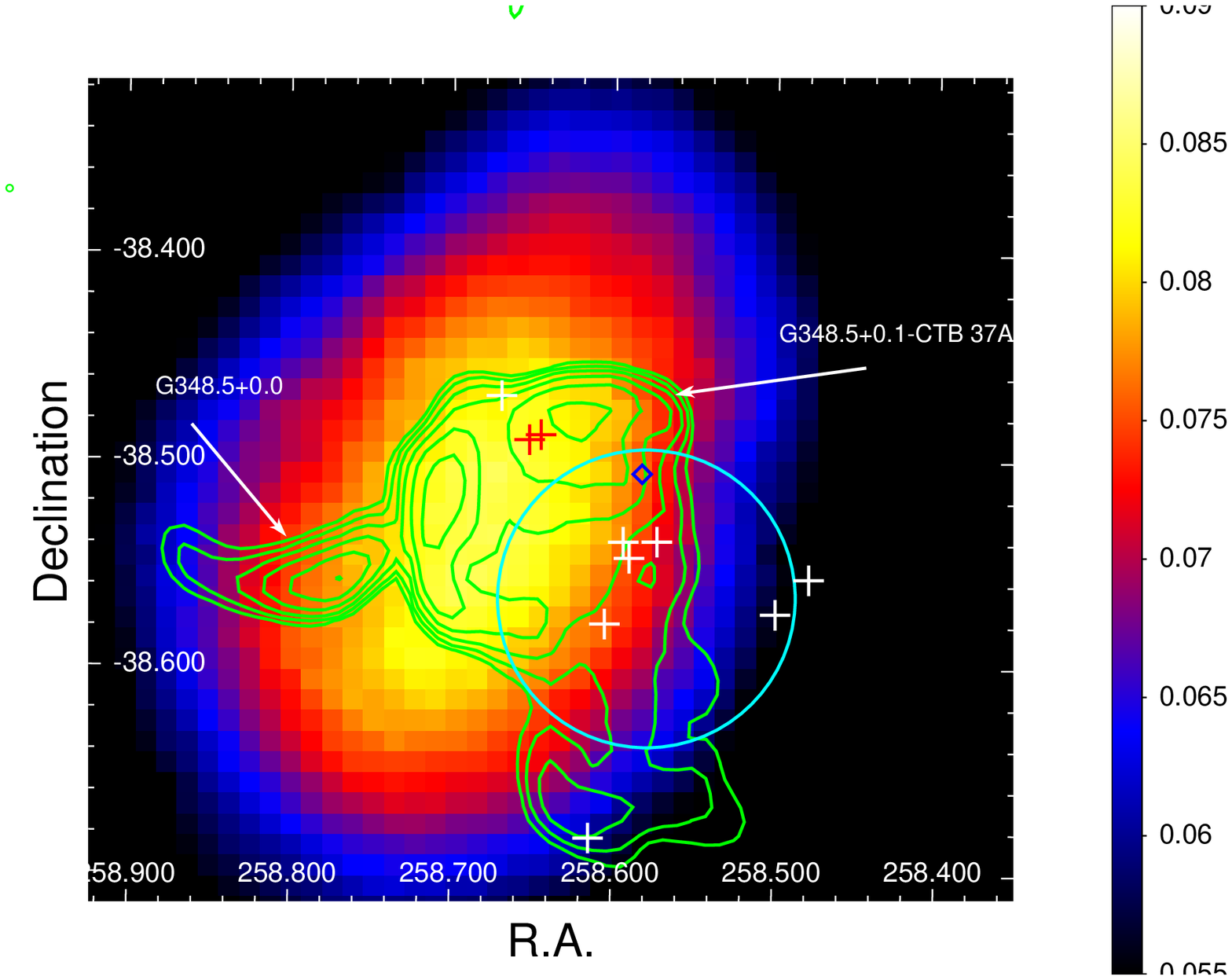}  \\
\includegraphics[width=0.45\textwidth]{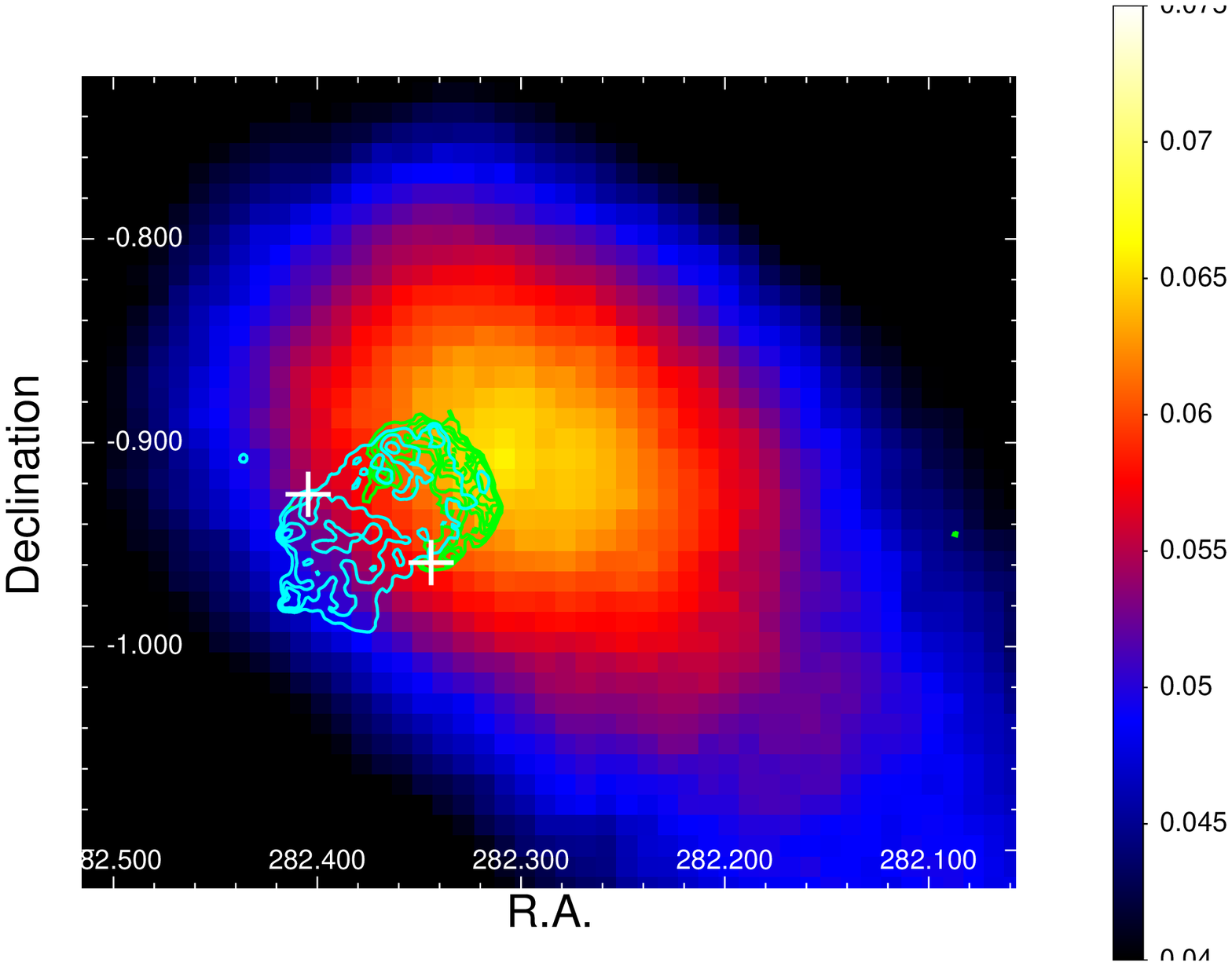}  
\includegraphics[width=0.45\textwidth]{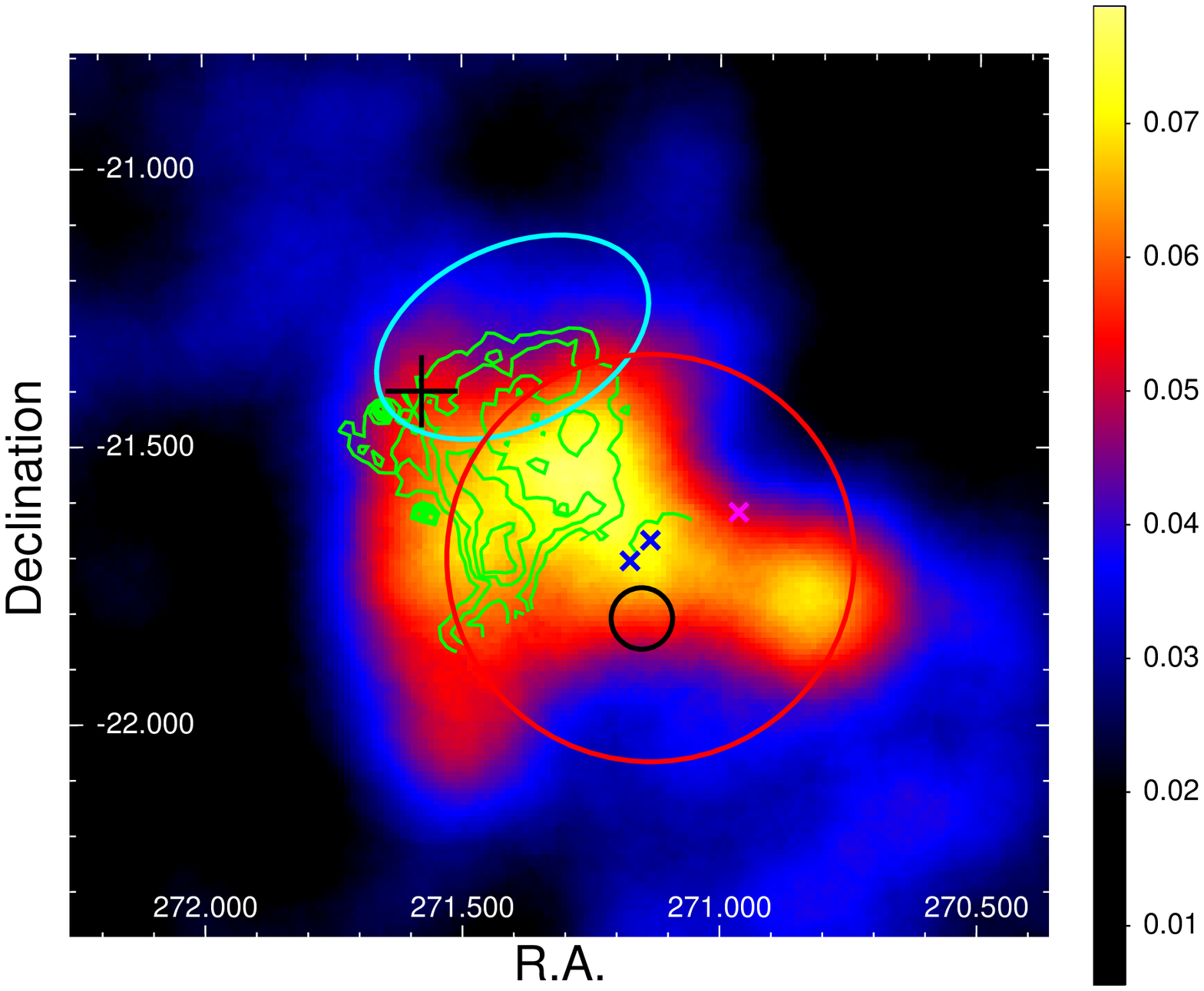}  
\caption{\footnotesize{Smoothed {\it Fermi}-LAT count maps of {\it front} converted events in the range 2-200 GeV (units are $10^4$ $\text{counts deg}^{-2}$), of the regions surrounding the remnants studied. The pixel binning is 0$^\circ$.01, and the maps are smoothed with Gaussians of width 0$^\circ$.2. In (a) {\it Chandra} X-ray emission map contours are overlaid. Black crosses indicate the direction of the detected OH (1720 MHz) maser emission. For CTB 37A, in (b), radio emission, as observed by \citet{Whiteoak1996} in the Molonglo plane survey at 843 MHz, is overlaid as green contours. White crosses indicate the direction of the detected OH (1720 MHz) maser emission at velocities $\sim -65$ km s$^{-1}$ associated with CTB 37A, and the red crosses show the positions of maser emission at $\sim  -22$ km s$^{-1}$, associated with G348.5+0.0. The blue diamond indicates the position of CXOU J171419.8-383023, and the cyan circle shows the 1$\sigma$ containment radius of HESS J1714-385. In (c), the green contours show the 20 cm emission toward 3C 391, as seen by the VLA FIRST Galactic Plane Survey, see \citet{White2005}. Additionally, the cyan contours show the {\it Chandra} X-ray emission map. Yellow crosses indicate the direction of the two detected OH (1720 MHz) maser emission spots associated with 3C 391. In (d), the green contours show the diffuse 20 cm emission toward G8.7--0.1, as seen by the VLA FIRST Galactic Plane Survey. The black cross indicates the direction of the detected OH (1720 MHz) maser emission.The cyan ellipse contains the area where soft X-ray emission (0.1-2.4 keV) is observed with {\it ROSAT}, and the red circle indicates the approximate extension of HESS J1804-216.The black circle contains the radio extension of SNR G8.31-0.09, recently detected by \citet{Brogan2006}. The blue crosses indicate the position of {\it Chandra} sources CXOU J180432.4-214009 and CXOU J180441.9-214224 \citep{Bamba2007, Kargaltsev2007b}. The magenta cross indicates the position of the young Vela-like pulsar B1800-21, found by \citet{Clifton1986,Kargaltsev2007a}.}
}
\label{fig:im}
\end{center}
\end{figure*}
\end{center}

G349+0.2 has one of the highest radio and X-ray surface brightnesses in the Milky Way. Its radio image is characterized by a small diameter (approximately $2'.5$) partial shell,  somewhat enhanced toward the southern and eastern limbs \citep{Shaver1985}. Five OH maser emission spots (at 1720 MHz) were detected in the direction of this SNR, which suggests that it is interacting with a dense molecular cloud \citep{Frail1996}. The velocity of these maser lines places the regions of shocked material at a distance $d\simeq 22.4$ kpc. The characteristics of the emission detected from molecular transitions of $^{12}$CO, $^{13}$CO, CS, HCO$^{+}$, H$_{2}$CO, HCN and SO, at similar velocities to those of the OH masers, are further evidence of molecular cloud-shock interaction \citep{Lazendic2004,Reynoso2001}.

\begin{center}
\begin{figure*}
\begin{center}
\includegraphics[width=0.45\textwidth]{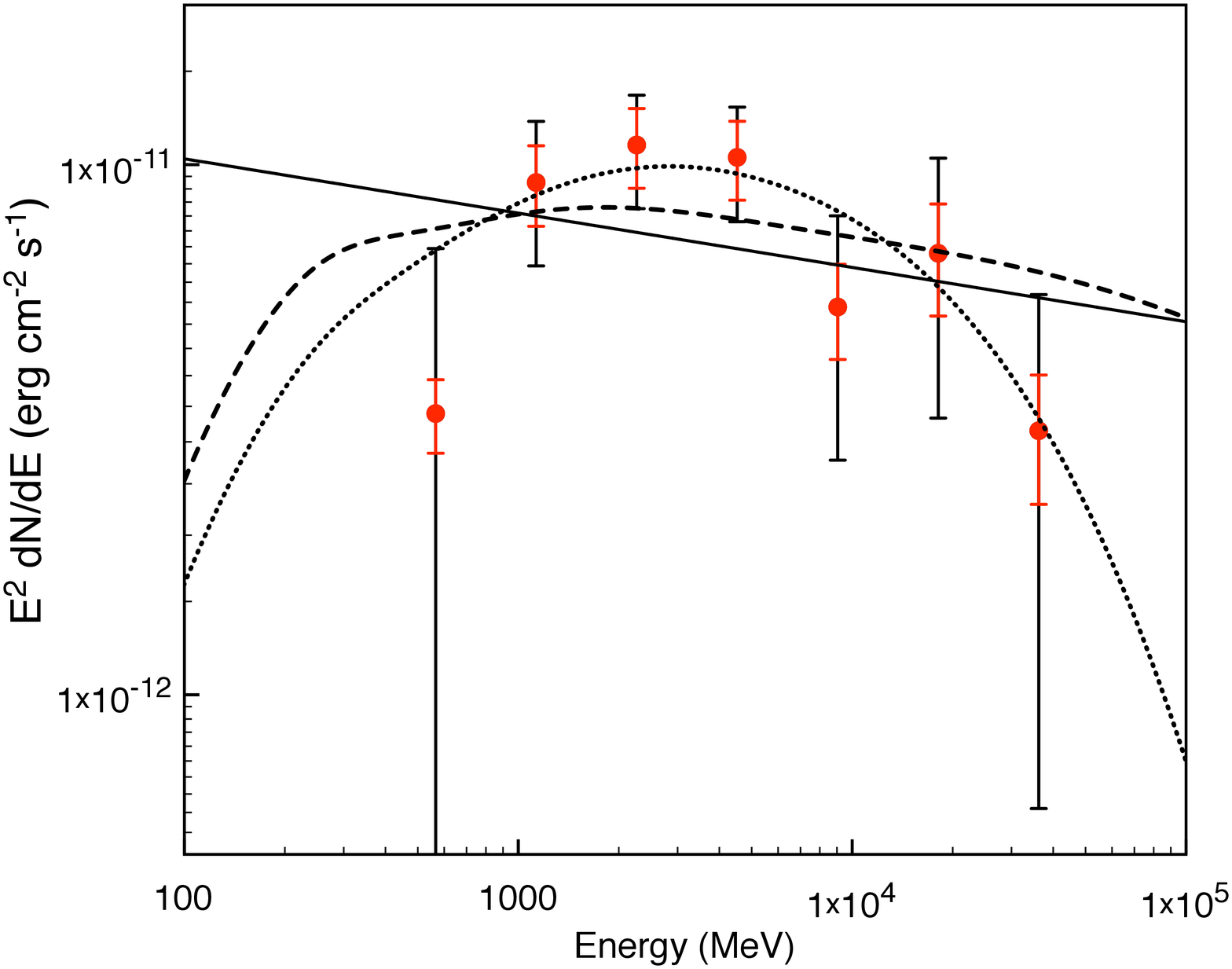}  
\includegraphics[width=0.45\textwidth]{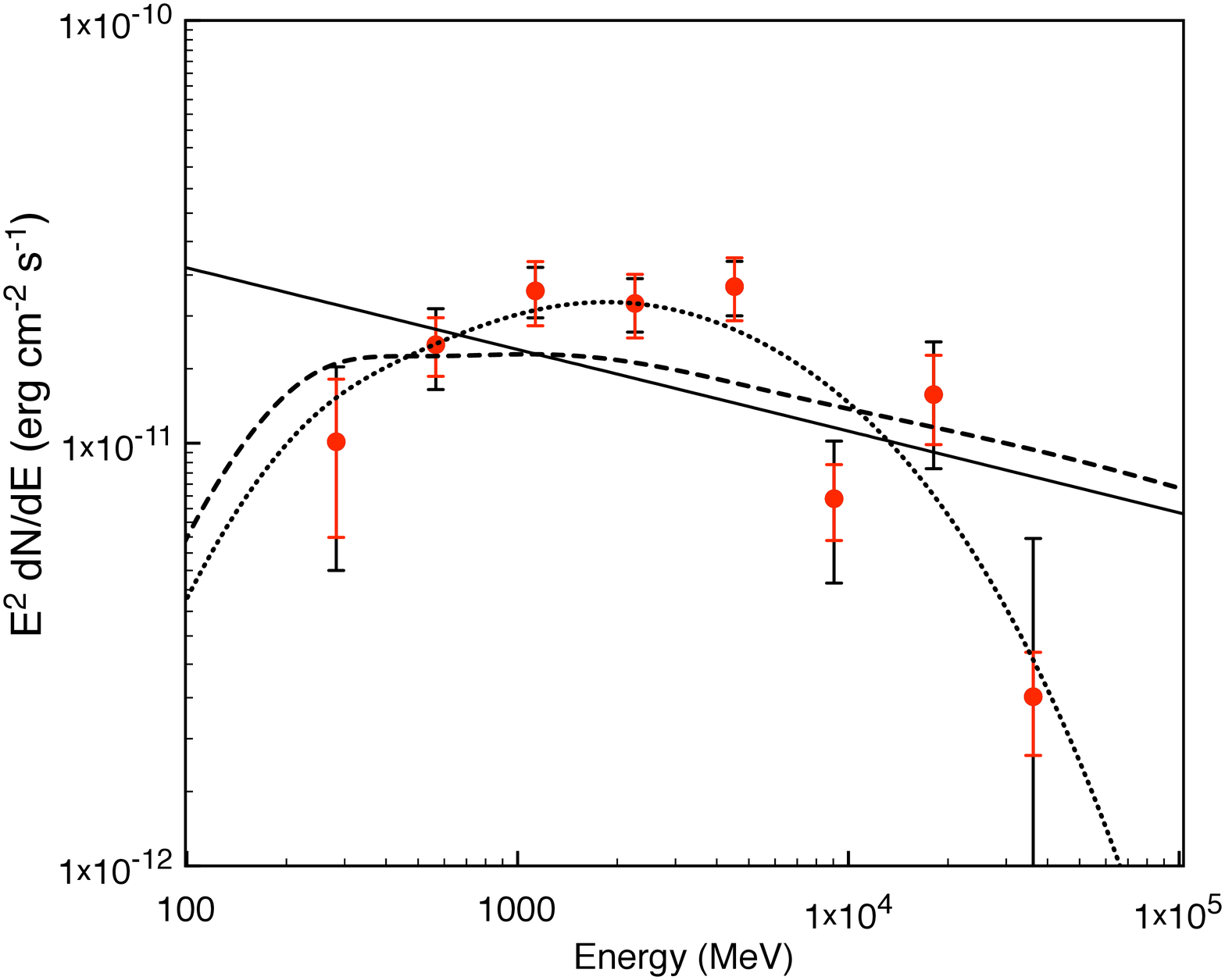}  \\
\includegraphics[width=0.45\textwidth]{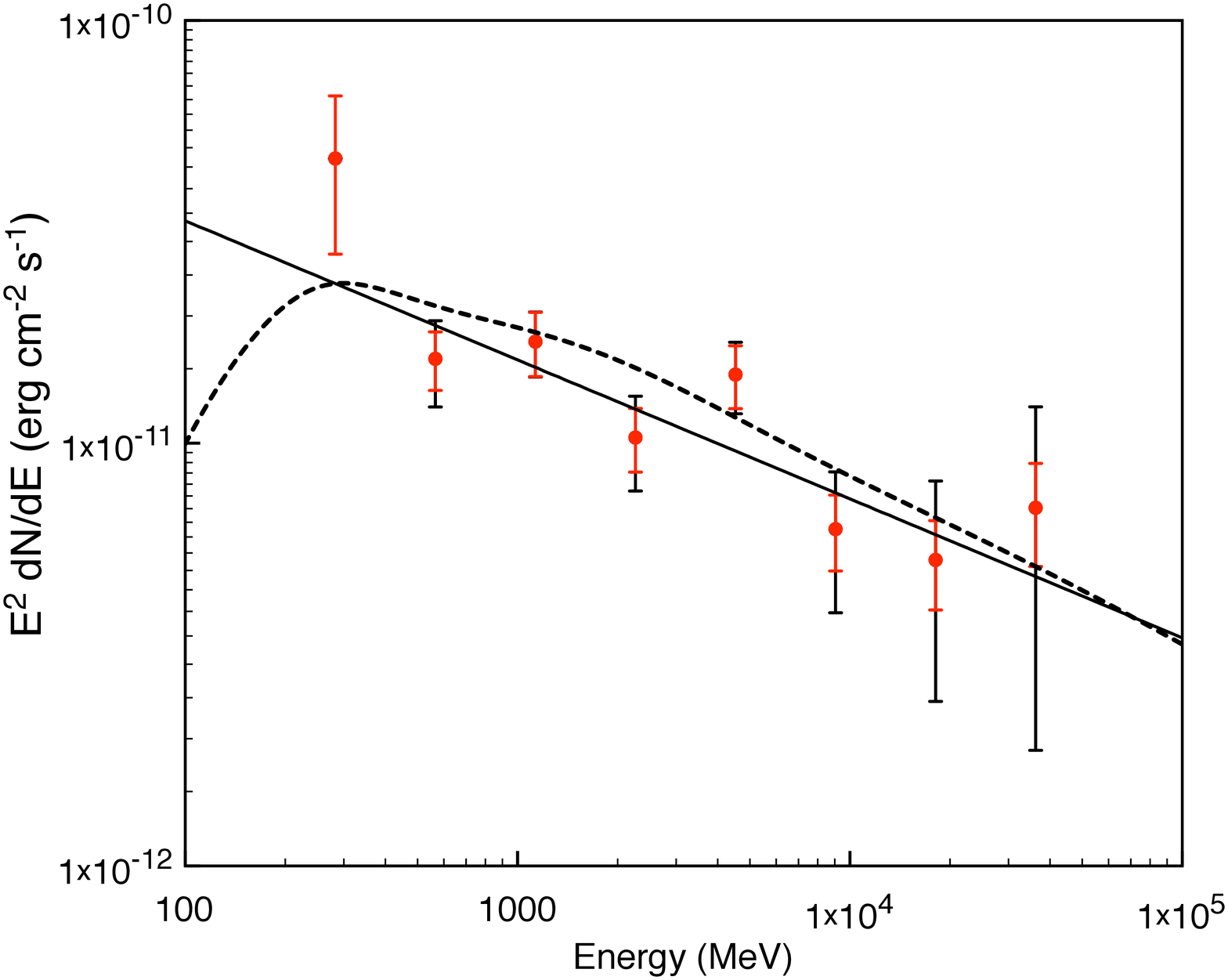}  
\includegraphics[width=0.45\textwidth]{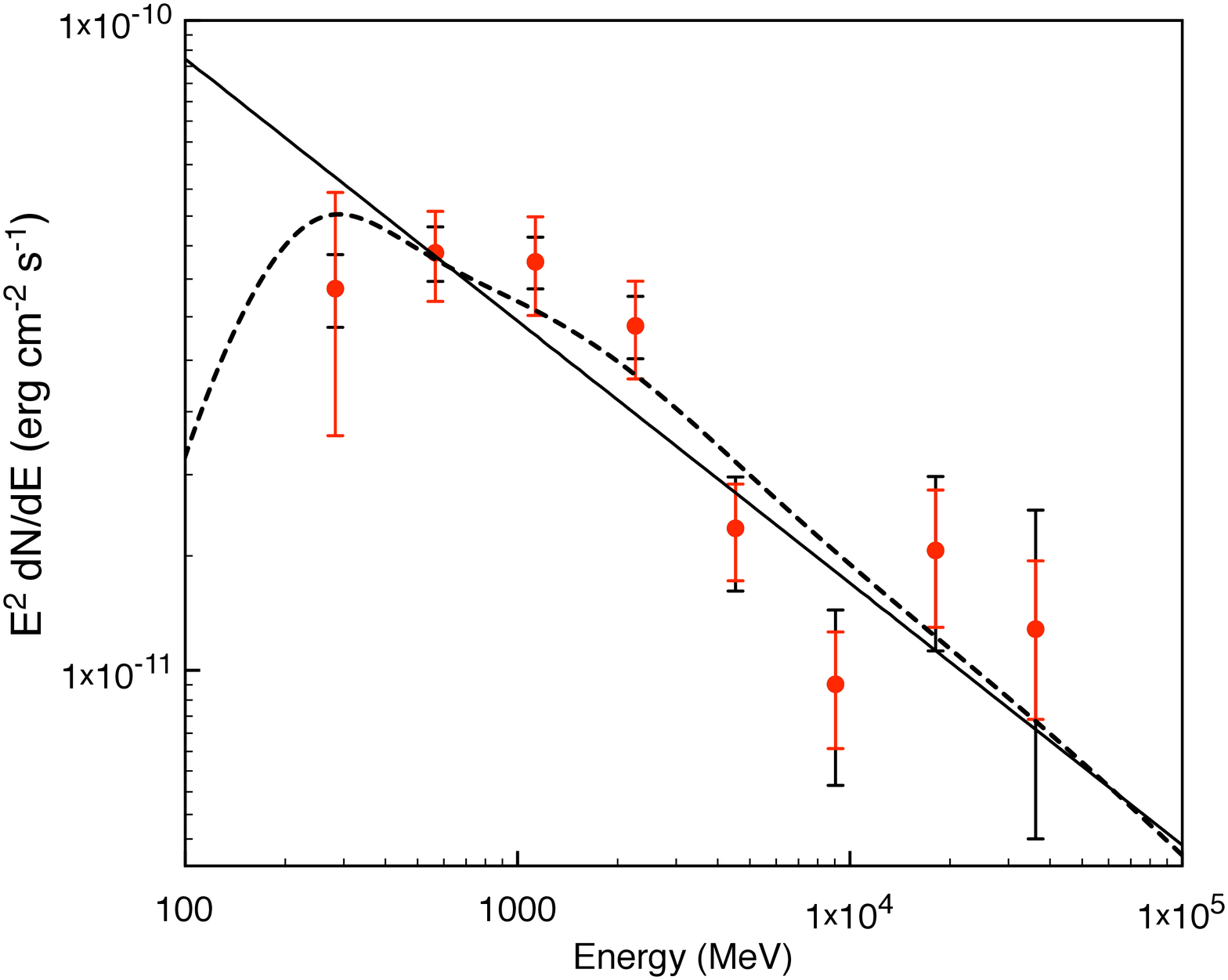}  
\caption{\footnotesize{{\it Fermi}-LAT SED of the sources coincident with SNRs (a) G349.7+0.2, (b) CTB~37A, (c) 3C~391, and (d) G8.7--0.1. Statistical uncertainties are shown as black error bars, and systematic errors are indicated by red bars. The best-fit power-law models are shown (solid black lines), as well as the spectra from $\pi^0$-decay models for proton distributions with energy cutoff at 100 TeV (dashed curves). For G349.7+0.2 and CTB 37A the spectra resulting from lower proton energy cutoffs, 160 GeV and 80 GeV respectively, are also shown (dotted curves).}
}
\label{fig:spec}
\end{center}
\end{figure*}
\end{center}

In the X-ray band, the morphology of the remnant observed with {\it ASCA} and {\it Chandra} is similar to that from the radio, and the high photon absorption of its X-ray spectrum ($N_\text{H} \approx 7\times10^{22}$ cm$^{-2}$) is consistent with the distance estimate from maser emission \citep{Yamauchi1998,Slane2002,Lazendic2005}. In the analysis of the {\it Chandra} observations, \citet{Lazendic2005} find that the overall X-ray emission of the remnant is best-fit by a nonequilibrium ionization plasma model with enhanced Si abundance, and temperature $kT\approx1.4$ keV. The fit is significantly improved with the addition of a softer component of equilibrium thermal plasma with solar abundances and temperature $kT\approx0.8$ keV. The Si overabundance suggests that the harder spectral component is emission from the supernova ejecta, and the softer component, with solar abundances, is related to shocked circumstellar material. From the volume emission measure these authors estimate the hydrogen density for the soft component to be $n_s\gtrsim 4.2f_V^{-1/2}$ cm$^{-3}$, where $f_V$ is the volume filling factor, and assuming that the distance is 22 kpc and $n_e\approx1.2n_{\text{H}}$. The {\it Chandra} observations also revealed X-ray emission from a point source, CXOU J171801.0-372617, inside the SNR shell, with X-ray characteristics similar to the compact central object (CCO) class, such as the source in Cas A, and unlike those for energetic pulsars \citep{Pavlov2004}.

The {\it Fermi} LAT count map around G349.7+0.2 is shown in Figure \ref{fig:im}a, overlaid with the contours of the {\it Chandra} X-ray map. A bright unresolved $\gamma$-ray source is coincident with the position of the remnant, and with that of the source designated as 1FGL J1717.9-3729 in the one-year {\it Fermi} LAT First Source Catalog. The significance of the detection, obtained from the evaluation of the test statistic in the region, is $\sim 10\sigma$.  The residual test statistic map shows no evidence that the source is spatially extended. The spectrum of the $\gamma$-ray emission is shown in Figure \ref{fig:spec}a, and its shape suggests that the spectral index steepens above a few GeV. The best-fit model for the data yields a spectral index of $\Gamma=-2.10\pm0.11$. The addition of an exponential cutoff is found to marginally improve the fit, the cutoff energy is estimated to be $\sim 16.5$ GeV, and the spectral index is $\Gamma=-1.74\pm0.37$. Table \ref{tab:specs} summarizes the characteristics of the best fit spectral model of the emission from this source, as well as its position. The luminosity in the 0.1-100 GeV band of this source is estimated to be $\sim 1.3\times10^{35}$erg s$^{-1}$, at a presumed distance of 22 kpc.

\subsection{CTB 37A}

The SNR CTB 37A (G348.5+0.1) is located in a complicated region composed of three different remnants (Figure \ref{fig:im}b). Its radio emission overlaps that of a second SNR, G348.5-0.0 on the east \citep{Kassim1991}, and a third remnant, CTB 37B, lies approximately  20$'$ to the north. The radio morphology of CTB 37A appears as a well-defined partial shell towards the north and east, and an extended outbreak to the south. H{\footnotesize \,I} absorption measurements by \citet{Caswell1975}, toward both CTB 37A and CTB 37B, place constraints on the distance to the CTB 37A/B complex, which is estimated to lie between 6.6 kpc and 11.4 kpc, using a distance to the Galactic center of 8.5 kpc.

\begin{table*}
\begin{center}
\caption{\footnotesize{Summary of spatial and spectral fit parameters from the {\it Fermi}-LAT data}}
\label{tab:specs}
\begin{tabular}{lcccccccccc}
\noalign{\smallskip}
\toprule
\noalign{\smallskip}
&&\multicolumn{3}{c}{Spatial}& &\multicolumn{3}{c}{Spectral Fit}&& \\ 
\noalign{\smallskip}
\cline{3-5} \cline{7-9}
\noalign{\smallskip}
\noalign{\smallskip}
& &R.A. & Decl. &$r_{95}$& & {\it F}(0.1-100 GeV) & &&&\\
Name & &(deg)&(deg)&(deg)&&(10$^{-7}$ photons cm$^{-2}$ s$^{-1}$)& $\Gamma$ & $\chi^2$(dof)&&TS Value\\
\midrule
G349.7+0.2 &&259.47&-37.50&0.03&&$0.58 \pm 0.11$&$2.10 \pm 0.11$&1.12(5)&&108\\
\noalign{\smallskip}
CTB 37A&&258.65&38.52&0.04&&$1.36 \pm 0.15$&$2.19 \pm 0.07$&2.64(6)&&294\\
\noalign{\smallskip}
3C 391&&282.26&-0.92&0.03&&$1.58 \pm 0.26$&$2.33 \pm 0.11$&0.80(6)&&140\\
\noalign{\smallskip}
G8.7--0.1&&271.33&-21.64&0.03&&$3.88 \pm 0.42$&$2.40 \pm 0.07$&1.22(6)&&486\\
\bottomrule
\end{tabular}
\end{center}
\end{table*}

Various OH masers have been detected toward the interior of CTB 37A at 1720 MHz \citep{Frail1996}. Eight of these have velocities close to the upper velocity limit for this object given by the neutral hydrogen measurements (-65 km s$^{-1}$). At this velocity, \citet{Reynoso2000} observed, in the CO ($J=1\rightarrow 0$) transition at 115 GHz, three molecular clouds coincident with the maser spots and estimated the distance to the clouds to be 11.3 kpc. 

\citet{Aharonian2008}  detected extended thermal X-ray emission from inside the NE shell of the remnant, using observations with {\it Chandra} and {\it XMM-Newton}. This study found consistent fit results from the detectors on both X-ray observatories, with plasma temperature T$\sim0.8$ keV and absorption $N_\text{H} \sim 3\times10^{22}$ cm$^{-2}$. Additionally, these authors observed an extended non-thermal X-ray source, CXOU J171419.8-383023, toward the NW of the CTB 37A complex (see Figure \ref{fig:im}b), which they suggest might be associated with the emission of a pulsar wind nebula (PWN). A VHE ($E>$100 GeV) $\gamma$-ray source, HESS J1714-385, is coincident with CTB 37A (Figure \ref{fig:im}b), and it has been suggested this is a signature of the interaction between the remnant's shock wave and the surrounding clouds of dense material \citep{Aharonian2008}.

In Figure \ref{fig:im}b, the smoothed GeV map of the region around CTB 37A is shown. The $\gamma$-ray emission in the {\it Fermi} LAT range appears as an unresolved source whose centroid lies within the eastern radio shell of CTB 37A, with detection significance 17$\sigma$. This source is coincident with 1FGL J1714.5-3830. The residual test statistic map gives no indication that this source is extended beyond the PSF of the instrument. The spectrum of this source is displayed in Figure \ref{fig:spec}b, together with the best-fit models. The simple power-law model for the data yields a spectral index $\Gamma=-2.19\pm0.07$. However, the fit is slightly improved by including an exponential cutoff at $E_{\text{cut}}\sim4.2$ GeV, with spectral index $\Gamma=-1.46\pm0.32$. Table \ref{tab:specs} summarizes the result of the analysis of this source. The luminosity in the 0.1-100 GeV band of this source is estimated to be $\sim 1.3\times10^{35}$erg s$^{-1}$, at a presumed distance of 11.3 kpc.

\subsection{3C 391}	

3C391 (G31.9+0.0) is a radio bright SNR with an irregular spatial structure. Observations with the Very Large Array (VLA)\footnote{The VLA is operated by the National Radio Astronomy Observatory, which is a facility 
of the National Science Foundation, operated under cooperative agreement by Associated 
Universities, Inc.} in the radio band uncovered an elongated morphology, which extends from northwest to southeast \citep{Reynolds1993}. The shell surrounding the radio emission is enhanced toward the NW, and does not extend to the SE, where it appears that the SNR has broken out into a lower density region. 

\citet{Frail1996} observed two OH maser spots coincident with 3C 391 at 1720 MHz, with velocities (105 and 110 km s$^{-1}$) that agree well with the neutral hydrogen absorption velocity \citep{Radhakrishnan1972}.  The H{\footnotesize \,I} absorption measurements indicate a distance of at least 7.2 kpc, and the absence of absorption at negative velocities suggests an upper limit of 11.4 kpc, using a distance to the center of the Galaxy of 8.5 kpc. The case for the interaction of 3C 391 with a molecular cloud, as suggested by the maser emission in the direction of the SNR, is further strengthened by the observations of CO ($J=1\rightarrow 0$) emission in the region \citep{Wilner1998}. The CO maps show evidence that 3C 391 resides at the edge of a molecular cloud located to the NW of the SNR, and that the "breakout" morphology is produced as the SNR blast wave reaches lower density regions outside the cloud. 

Observations with {\it ASCA} \citep{Chen2001} and {\it Chandra} \citep{Chen2004} show that the X-ray emission associated with this object is thermal and clumpy in nature, and its brightness peaks well inside the radio shell. The plasma temperature derived from the X-ray observations is $\sim0.5-0.6$ keV, and the absorption column density is found to increase from SE to NW (from 0.6 to 4.1 $\times 10^{22} \text{cm}^{-2}$), which is consistent with the existence of molecular cloud to the northwest \citep{Chen2004}. These authors also discovered an unresolved X-ray source on the NW border, with spectral characteristics similar to those of CCO class of objects.

The smoothed photon count map from the {\it Fermi}-LAT data is shown in Figure \ref{fig:im}c, and presents a strong case for an association between a GeV source and 3C 391. The unresolved $\gamma$-ray source is detected with a 13$\sigma$ significance and the peak of the test statistic map is 4$'$ from the northwestern radio shell and the position of the closest maser spot. This source is coincident with 1FGL J1849.0-0055. The SED of the emission from this source is best described by a power law model, with spectral index $\Gamma=-2.33\pm0.11$. The SED is shown in Figure \ref{fig:spec}c, and spectral characteristics are presented in Table \ref{tab:specs}. The luminosity of this source in the 0.1-100 GeV band is $\sim 0.9\times10^{35}$erg s$^{-1}$, at a presumed distance of 8 kpc.

\subsection{G8.7--0.1 (W30)}

The rather large SNR G8.7--0.1, together with nine discrete H{\footnotesize \,II} regions, forms the W30 complex. G8.7--0.1 was identified as a Galactic SNR on the basis of the observation of a 45$'$ diameter area of extended non-thermal radio emission \citep{Odegard1986,Kassim1990}. 

\citet{Hewitt2009a}  detected a single OH maser at 36 km s$^{-1}$, on the eastern edge of G8.7--0.1. This relatively bright emission at 1720 MHz does not coincide with any of the compact radio sources, and hence they associate it with the SNR, and using the rotation curve of \citet{Fich1989} they estimate a distance to the remnant of $4.5$ kpc. An extended $\gamma$-ray source, HESS J1804-216, has been detected to the southwest of the remnant (Figure \ref{fig:im}d), which could be further evidence of shock interaction with clouds of dense material \citep{Aharonian2006}. Several other sources in the field have been suggested as possible associations with the source HESS J1804-216, and they are shown in Figure \ref{fig:im}d \citep{Clifton1986, Brogan2006, Bamba2007, Kargaltsev2007b,Kargaltsev2007a,Higashi2008}.

\citet{Finley1994} detected diffuse thermal X-ray emission from the northern half of the remnant in the 0.1-2.4 keV band with {\it ROSAT}, and using Sedov analysis estimated the distance to the remnant to be $3.2$ kpc $\leq d\leq 4.3$ kpc. The estimated plasma temperature from these X-ray observations is $T\approx 0.3-0.7$ keV, the absorption column density is $N_\text{H} \sim 1.2-1.4\times10^{22}$ cm$^{-2}$, and the derived ambient density is $n_0 \sim 0.02-0.04 \text{ cm}^{-3}$. These authors also determined the distance to the H{\footnotesize \,II} regions associated with remnant to be $\sim4.8$ kpc, using the kinematic measurements of \citet{Kassim1990} and more recent Galactic rotation models.

The $\gamma$-ray photon count map in the 2-200 GeV range of the crowded vicinity of the W30 complex is shown in Figure \ref{fig:im}d. The $\gamma$-ray emission in the region appears as a source centered on a position inside the radio boundary of G8.7--0.1. This detection is coincident with 1FGL J1805.2-2137. The residual test statistic map does not rule out the possibility that  this source is extended. The significance of the detection is $\sim 22\sigma$. The spectrum of this source, shown in Figure \ref{fig:spec}d, seems to steepen above 2 GeV, and flatten at $\gtrsim 10$ GeV. For a single power-law model the best-fit spectral index is $\Gamma=-2.40\pm0.07$. The spectral characteristics of this region are summarized in Table \ref{tab:specs} . The luminosity in the 0.1-100 GeV band of this source is $\sim 1.1\times10^{35}$erg s$^{-1}$, at a presumed distance of 4.5 kpc.

\section{DISCUSSION}

The detection with the {\it Fermi}-LAT of unresolved $\gamma$-ray sources in the direction of G349.7+0.2, CTB 37A, and 3C 391, as well as a possibly extended region of emission coincident with G8.7--0.1 is reported in this work. The best-fit positions and spectral characteristics of these sources are summarized in Table \ref{tab:specs}.

The shapes of the spectra of the sources coincident with G349.7-0.5, 3C 391 and G8.7--0.1 do not seem consistent with what has been observed from pulsars in $\gamma$-rays. \citet{Abdo2009b} found the SED of pulsars in the {\it Fermi} LAT band to be best described by power-law distributions, with exponential cutoffs in the energy range 1-6 GeV (the only exception, PSR J1833-1034 with cutoff energy $E_{\text{cut}} \sim$ 8 GeV, was reported to be poorly measured). The spectral characteristics of the sources coincident with 3C 391 and G8.7--0.1 do not 
suggest an exponential cutoff at all, and the spectrum of G349.7+0.2 is very modestly improved by a cutoff at 
$E_{\text{cut}}=16.5\pm1.9$ GeV, which is very high for a $\gamma$-ray pulsar. Additionally, there are no pulsars from the Australian Telescope National Facility (ATNF) Pulsar Catalogue\footnote{The positions and characteristics of the 1509 pulsars in ATNF catalogue, as well as further documentation, can be accessed at http://www.atnf.csiro.au/research/pulsar/psrcat.} coincident with this source \citep{Manchester2005}.

The spectrum of the source coincident with CTB 37A is well fit by a power law with an exponential cutoff at energy $E_{\text{cut}} \sim4.2$ GeV, and hence does not rule out a pulsar hypothesis. However, there are no ATNF pulsars within 20$'$ radius of this source. 

Given the lack of evidence for significant contribution by a pulsar, and the known presence of maser emission for each remnant, an appropriate hypothesis appears to be that the $\gamma$-ray sources are
products of SNR interactions with molecular clouds. To investigate properties of the expected spectra, we have considered the scenario in which the $\gamma$-rays are produced by $\pi^0$ decay of accelerated
hadrons. In order to reproduce the observed spectra, we have used a model based on that from \citet{Kamae2006} using a scaling factor of 1.85 for helium and heavier nuclei \citep{Mori2009}, and adopted
a power law model with an exponential cutoff for the energy distribution of the protons:

\begin{equation}
\frac{dN_p}{dE_p} = a_p \left( \frac{E_p}{\text{1 TeV}}\right)^{-\Gamma_p} \exp\left[-\frac{E_p}{E_0}\right],
\label{eq:uno}
\end{equation}

where $E_0$ is the proton energy cutoff. Table \ref{tab:dens} presents sets of representative model parameters that adequately reproduce the observed $\gamma$-ray spectra (see Figure 2). While the spectra of SNRs 3C 391 and G8.7--0.1 are well matched by models with proton distributions with spectral indices $\Gamma\approx 2.4$ and high-energy cutoff, $E_0>100\text{ TeV}$, the spectra of G349.7+0.2 and CTB 37A  require harder proton distributions, $\Gamma\approx1.7$, and $E_0\sim 10^2 \text{ GeV}$. The model provides an estimate of the density of the material the SNR shock wave is impacting, given reasonable assumptions about the shock compression ratio, $r$, and the fraction $\epsilon$ of the total supernova explosion energy $E$ that is converted into cosmic ray energy. Here we assume $r = 4$, $\epsilon = 0.4$, and $E = 10^{51}$~erg; we use distances derived from observations at other wavelengths for each SNR. In Table {\ref{tab:dens}}, estimates of the densities of the $\gamma$-ray emitting material are compared to the densities derived from X-ray observations in other studies \citep{Lazendic2005,Chen2004,Finley1994}. 

\begin{table}
\begin{center}
\caption{\footnotesize{Model parameters, distances, and density estimates}}
\label{tab:dens}
\begin{tabular}{lcccccr}
\toprule
&&\multicolumn{3}{c}{Model Parameters}& &\\ 
\noalign{\smallskip}
\cline{3-5}
\noalign{\smallskip}

Object &&$\Gamma_p$&$E_0$&{\it n}$_{\gamma}^{\,\,\,\,\text{a}}$& {\it n}$_{\text{X}}^{\,\,\,\,\text{b}}$&\it d  \\
  &&&(TeV)&(cm$^{-3})$&(cm$^{-3})$ &  (kpc)\\
\midrule
G349.7+0.2&&1.7&0.16&65&2.5&22\\
\noalign{\smallskip}
CTB 37A&&1.7&0.08&37&...&11.3\\
\noalign{\smallskip}
3C 391&&2.4&100&28&1&8\\
\noalign{\smallskip}
G8.7-0.1&&2.45&100&18&0.03&4.5\\
\bottomrule
\noalign{\smallskip}
\multicolumn{7}{p{0.9\columnwidth}}{\footnotesize{$^{\text{a}}$Densities derived using the $\gamma$-ray spectrum obtained from the analysis of {\it Fermi}-LAT observations.}}\\
\multicolumn{7}{p{0.9\columnwidth}}{\footnotesize{$^{\text{b}}$Densities derived from X-ray observations (see text for references). No estimate of the ambient density of SNR CTB 37A is provided in Aharonian et al. (2008).}}\\
\end{tabular}
\end{center}
\end{table}
%

The values obtained from the $\pi^0$-decay model fits to the $\gamma$-ray spectra of the SNRs are much higher than those from X-ray observations. It is possible that, upon interacting with dense ambient material, instabilities in the postshock flow result in considerable clumping of material. The denser regions would then not be expected to yield significant X-ray emission, although radiative shocks would then be expected to produce optical emission from these regions. If the filling fraction of the clumped material is large, then the
densities inferred for the X-ray emitting material are underestimates, potentially leading to more reasonable values for the density contrast between clump and interclump regions. 

Alternatively, the energetic particles that escape the particle acceleration region at the SNR shock could stream ahead of the shock and encounter clouds of molecular material, where the much higher density could dramatically enhance the $\gamma$-ray emission \citep{Aharonian1996,Gabici2007}. This scenario has been proposed as a possible explanation of the observed $\gamma$-ray fluxes in the direction of some Galactic SNRs, including RX J1713.7-3946 \citep{Butt2001} and W28 \citep{Fujita2009}. \citet{Lee2008} model the interaction of particles accelerated by the diffusive shock of an SNR with arbitrary matter distributions outside the remnant blast wave, including an appropriate treatment of the escape and diffusion of these particles. These authors, as well as other works \citep{Gabici2007,Rodriguez2008,Gabici2008}, show that clouds of dense material in close proximity to SNRs can have an important, and even dominant contribution to the $\gamma$-ray flux from these regions. It is important to note, however, that the escaping particles are predominantly from the high-energy portion of the spectrum; the observed spectra, which extend to low energies, may be problematic for such a scenario.  

An alternative scenario is that  the $\gamma$-ray emission coincident with these sources is of leptonic origin. Bremsstrahlung and inverse Compton (IC) radiation from the nonthermal electron population have been proposed as production mechanisms of $\gamma$-rays in SNRs \citep[e.g.][]{Bykov2000}, but these models present certain difficulties as well. Non-thermal bremsstrahlung can be the dominant emission process in this band for electron-to-proton ratios, $a_e/a_p$, greater than $\sim 0.2$ \citep{Gaisser1998}. However, cosmic-ray abundance ratios suggest $a_e/a_p\sim 10^{-2}$, and models for $\gamma$-ray emission from other SNRs (e.g., RX J1713.7-3946) favor even smaller values \citep{Morlino2009,Zira2010,Ellison2010}, making the bremsstrahlung-dominated model appear less likely. The IC-dominated scenario, on the other hand, is difficult to reconcile with the observed characteristics of the sources studied in this work. The expected spectral index for IC emission ($\Gamma \approx 1.5$) is considerably harder than the values obtained here ($\Gamma > 1.7$). 

In summary, using observations with the {\it Fermi}-LAT, we have revealed $\gamma$-ray emission from four SNRs known to be interacting with molecular clouds. In each case the inferred density is high, assuming a hadronic origin for the $\gamma$-rays, consistent with the presence of dense material indicated by maser emission. However, estimated densities of the ambient material derived under simple geometric assumptions are in considerable excess of those derived from the hot X-ray gas observed in the SNRs. More detailed modeling of the X-ray and $\gamma$-ray emission under a variety of scenarios for the ambient medium is required to address the emission mechanism more fully. These results are consistent with the scenario presented by \citet{Abdo2009a} for SNR W51C, and further studies of other such systems and their GeV-scale $\gamma$-ray emission with the {\it Fermi}-LAT are of particular interest.

\acknowledgments
The authors thank Yasunobu Uchiyama and Stefan Funk for their careful review of this work, and Don Ellison, Keith Bechtol, Joshua Lande, Andrei Bykov, Luke Drury, and Felix Aharonian for helpful discussions. This work was partially funded by NASA contract NAS 8-03060 and Fermi grant NNX09AT68G. P.O.S. acknowledges the KITP in Santa Barbara, where work on this project was begun while participating in a KITP program.

\end{document}